\documentclass{article}

\usepackage[T1]{fontenc}
\usepackage{amssymb}
\usepackage{url}  
\usepackage{subfigure}
\usepackage{amsmath, amssymb}
\usepackage{amsfonts,amsmath,amssymb,epsfig,epsf,psfrag}
\usepackage[english]{babel}
\usepackage{enumerate}
\usepackage{url}
\usepackage{color}
\usepackage[utf8]{inputenc}
\usepackage{lmodern}
\usepackage{graphicx}
\usepackage[T1]{fontenc}
\usepackage{amsthm}
\usepackage{multimedia}
\usepackage{changebar}
\usepackage{natbib}

\newtheorem{theorem}{Theorem}[section]
\newtheorem{proposition}[theorem]{Proposition}
\newcommand{\pkg}[1]{{\normalfont\fontseries{b}\selectfont #1}}
\let\proglang=\textit
\let\code=\texttt

\definecolor{darkblue}{rgb}{0,0,0.545098} 
\definecolor{lightblue}{rgb}{0.6784314,0.8470588,0.9019608} 
\definecolor{blue}{rgb}{0,0,1} 
\definecolor{pink}{rgb}{1,0.7529412,0.7960784}
\definecolor{purple}{rgb}{0.6274510,0.1254902,0.9411765}

\begin{document}


\title{Segmentor3IsBack: an \proglang{R} package for the fast and exact segmentation of Seq-data}


\author{A. Cleynen$^{1}$; Michel Koskas$^{1}$; Emilie Lebarbier$^{1}$; Guillem Rigaill$^{2}$ and St\'ephane Robin$^{1}$}
\date{}
\maketitle \begin{center}\noindent $^{1}${\it AgroParisTech, UMR 518 MIA, 16, rue Claude Bernard, 75005 Paris, France\\
INRA, UMR 518 MIA, 16, rue Claude Bernard, 75005 Paris, France.} \\
\noindent $^{2}${\it URGV INRA-CNRS-Universit\'e d'\'Evry Val d'Essonne, 2 Rue Gaston Cr\'emieux,\\ 91057 Evry Cedex, France} \\
\end{center}


\maketitle

\begin{abstract}
\section*{Background:}
Genome annotation is an important issue in biology which has long been addressed with gene prediction methods and manual experiments requiring biological expertise. The expanding Next Generation Sequencing technologies and their enhanced precision allow a new approach to the domain: the segmentation of RNA-Seq data to determine gene boundaries.
\section*{Results:}
Because of its almost linear complexity, we propose to use the Pruned Dynamic Programming Algorithm, which performances had been acknowledged for CGH arrays, for Seq-experiment outputs. This requires the adaptation of the algorithm to the negative binomial distribution with which we model the data. We show that if the dispersion in the signal is known, the PDP algorithm can be used and we provide an estimator for this dispersion. We then propose to estimate the number of segments, which can be associated to coding or non-coding regions of the genome, using an oracle penalty.

\section*{Conclusions:}
We illustrate the results of our approach on a real data-set and show its good performance. Our algorithm is available as an \proglang{R} package on the CRAN repository.
\end{abstract}

\section*{Keywords}
segmentation algorithm, exact, fast, RNA-Seq data, count data


\section{Background}

Change-point detection methods have long been used in the analysis of
genetic data, for instance they proved a useful tool in the study of
DNA sequences with various purposes. \cite{braun_selection_1998, Lebarbier-Durot} have developed segmentation methods for categorical variables with the aim
of identifying patterns for gene predictions, while \cite{SNPsPattern}
uses the sequence segmentation for the detection of SNPs. In the last
two decades, with the large spread of micro-arrays, change-point
methods have been widely used for the analysis of DNA copy number
variations and the identification of amplification or deletion of
genomic regions in pathologies such as cancers \cite{Zhangseg,
  ErdmanEmerson2008, olshen_circular_2004,
  picard_statistical_2005,All2011}.

The recent development of Next-Generation Sequencing technologies
gives rise to new applications along with new difficulties: ($a$) the
increased size of profiles (up to $10^8$ data-points when micro-arrays
signals were closer to $10^5$), and ($b$) the discrete nature of the output
(number of reads starting at each position of the genome). Yet
applying segmentation methods to DNA-Seq data and its greater
resolution should lead to the analysis of copy-number variation with a
much improved precision than CGH arrays. Moreover, in the case of
poly-(A) RNA-Seq data on lower organisms, since coding regions of the
genome are well separated from non-coding regions with lower activity,
segmentation methods should allow the identification of transcribed
genes as well as address the issue of new transcript discovery.  Our
objective is therefore to develop a segmentation method to tackle
both ($a$) and ($b$) with some specific requirements: the amount of reads falling in a segment should be representative of the biological
information associated (relative copy-number of the region, relative
level of expression of the gene) and comparison to neighboring regions
should be sufficient to label the segment (for instance normal or
deleted region of the chromosome in DNA-Seq data, exon or non-coding
region in RNA-Seq), so that no comparison profile should be needed.
This also suppresses the need for normalization, and thus we wish to
analyze the raw count-profile.

So far, most methods addressing the analysis of these datasets require
some normalization process to allow the use of algorithms relying on
Gaussian-distributed data or previously developed for micro-arrays
\cite{Chiang,Tammi, yoon, Boeva}. Indeed, methods adapted to count
data-sets are not many, and highly focused on Poisson distribution. \cite{ShenZhang2012} proposes a method based on the comparison of Poisson processes associated with the read counts of a
case and a control sample, allowing for the detection of alteration of
genomic sequences but not for expressed genes in a normal
condition. \cite{rivera2012optimal} developed a likelihood ratio
statistic for the localization of a shift in the intensity of a
Poisson process while \cite{KirchPoisson} developed a test statistic
for the existence of a change-point in the Poisson autoregression of
order 1. Those two latter methods do not require a comparison profile
but they only allow for the detection of a single change-point and
have too high a time-complexity to be applied to RNA-Seq profiles. Binary Segmentation, a fast heuristic
\cite{olshen_circular_2004} and Pruned Exact Linear Time (PELT),
\cite{killick_pelt} an exact algorithm for optimal segmentation with
respect to the likelihood, are both implemented for the Poisson
distribution in package \pkg{changepoint}. Even though both are
extremely fast, do not require a comparison profile and analyse
count-data, the Poisson distribution is in-adapted to our kind of
data-sets.

A recent study of \cite{Hocking_short} has compared 13 segmentation
methods for the analysis of chromosomal copy number profiles and has
shown the excellent performances of the Pruned Dynamic Programming (PDP)
algorithm proposed by \cite{rigaill_pruned_2010} in its initial
implementation for the analysis of Gaussian data in the \proglang{R}
package \pkg{cghseg}. We propose to use the PDP algorithm which we have
implemented for the Poisson and negative binomial distributions.

In the next section we recall the general segmentation framework and the definition and requirements of the PDP algorithm. Our contributions are given in the third section where we define the negative binomial model and show that it satisfies the PDP algorithm requirements. We also give a model selection criterion with theoretical guaranties, which makes the whole approach complete. We conclude with a simulation study, which illustrates the performances of the proposed method.

\section{Segmentation model and algorithm}

\subsection{General segmentation model}

The general segmentation problem consists in partitioning a signal
of $n$ data-points $\{y_t\}_{t \in [\![1, n]\!]}$ into $K$ pieces or
segments. The model can be written as follows: the observed data
$\{y_t\}_{t=1,\ldots,n}$ are supposed to be a realization of an
independent random process $Y=\{Y_t\}_{t=1,\ldots,n}$. This process
is drawn from a probability distribution $\mathcal{G}$ which depends on a set
of parameters among which one parameter $\theta$ is assumed to be affected by
$K-1$ abrupt changes, called change-points, so that
$$
Y_t \sim \mathcal{G}(\theta_{r},\phi)  \qquad \mbox{if } t \ \in \ r \quad
\text{and} \quad r \in m
$$
where $m$ is a partition of  $[\![1, n]\!]$ into segments $r$, $\theta_r$ stands for the parameter of segment $r$ and $\phi$ is constant. The objective is to estimate the change-points or the positions of the
segments and the parameters $\theta_r$ both resulting from the segmentation. More precisely, we define $\mathcal{M}_{k,t}$ the set of all possible partitions in $k > 0 $ regions of the sequence up to point $t$. We remind that the number of possible partitions is
$$
\text{card}(\mathcal{M}_{K,t}) = {t-1\choose K-1}.
$$
We aim at choosing the partition in $\mathcal{M}_{K,n}$ of minimal loss
$\gamma$, where the loss is usually taken as the negative log-likelihood of the model. We define the loss of a segment with given parameter $\theta$ as $ c(r, \theta) = \sum_{i   \ \in \ r} \gamma(y_i, \theta)$, so  its optimal cost is  $c(r) =
\min_{ \theta} \left\{ c(r, \theta) \right\}$. This allows us to
define the cost of a segmentation $m$ as $\sum_{r \ \in
  \ m} c(r)$ and our goal is to recover the
optimal segmentation $M_{K,n}$ and its cost $C_{K,n}$ where :
\begin{eqnarray*}
M_{k,t} & = & {\arg\min}_{\{ m \ \in \ \mathcal{M}_{k,t} \}}
\left\{ \sum_{r \ \in \ m} c(r) \right\}  \\
\text{and} \quad C_{k,t} & = & {\min}_{\{ m \ \in \
\mathcal{M}_{k,t} \}} \left\{ \sum_{r \ \in \ m} c(r) \right\}.
\end{eqnarray*}

\subsection{Quick overview of the pruned DPA}

The pruned DPA relies on the function $H_{k,t}(\theta)$ which is the cost of the best  partition in $k$ regions up to $t$, the parameter of the last segment being $\theta$:
\begin{eqnarray*}
H_{k,t}(\theta) = \min_{k-1 \leq \tau \leq t} \{ \ C_{k-1,\tau} +
\ c([\tau+1, t], \theta) \ \},
\end{eqnarray*}
and from there gets $C_{k,t}$ as $min_{\theta} \{ H_{k,t}(\theta) \}$.  More
precisely, for each total number of regions, $k$, from $2$ to $K$, the
pruned DPA works on a list of last change-point  candidates:
$\text{ListCandidate}_k$. For each of these candidate change-points,
$\tau$, the algorithm stores a cost function and a set of optimal-cost
intervals.  To be more specific, we define:
\begin{itemize}
\item $ H_{k, t}^{\tau}(\theta) = C_{k, \tau} + \sum_{j=\tau+1}^{t}
  \gamma(y_{j}, \theta) $: the optimal cost if the last change
  is $\tau$;
\item $ S_{k, t}^{\tau} = \{ \theta \ | \ H_{k, t}^{\tau}(\theta) \ \leq
  \ H_{k, t}(\theta) \ \} $: the set of $\theta$ such that $\tau$ is
  optimal;
\item $ I_{k, t}^{\tau} = \{ \theta \ | \ H_{k, n}^{\tau}(\theta) \ \leq
  \ H_{k, n}^{t}(\theta) \ \}$: the set of $\theta$ such that $\tau$ is
  better than $t$ in terms of cost, with $\tau < t$.
\end{itemize}
We have $H_{k, t}(\theta) = \min_{\tau \leq t} \{H_{k, t}^{\tau}(\theta)\}$.

The PDP algorithm rely on four basic properties of these quantities:

\begin{enumerate}[$(i)$]
\item if all $\sum_{j=\tau+1}^{t+1} \gamma(y_{j}, \theta)$ are unimodal
  in $\theta$ then $I_{k, t}^{\tau}$ are intervals;
\item  $H_{k, t+1}^{\tau}(\theta)$ is obtained from $H_{k, t}^{\tau}(\theta)$ using:
\begin{eqnarray*}
H_{k,t+1}^\tau(\theta) = & H_{k,t}^\tau(\theta) + \gamma(y_{t+1}, \theta);
\end{eqnarray*}
\item it is easy to update $S_{k, t+1}^\tau$ using:
\begin{eqnarray*}
S_{k, t+1}^\tau & = & S_{k, t}^\tau \ \cap I_{k, t+1}^\tau \\
S_{k, t}^t & = & \complement_{\mathbb{R}} (\cup_{\tau \in [\![ k-1, t-1]\!]} I_{k, t}^\tau);
\end{eqnarray*}
\item once it has been determined that $S_{k, t}^\tau$ is empty, the
  region-border $\tau$ can be discarded from the list of candidates $ListCandidate_k$:
\begin{eqnarray*}
& S_{k, t}^\tau = \emptyset & \Rightarrow \qquad \forall \ t' \geq t
  \quad S_{k, t'}^\tau = \emptyset.
\end{eqnarray*}
\end{enumerate}

\paragraph{Requirements of the pruned dynamic programming algorithm.}
\begin{proposition}
Properties ($i$) to ($iv$) are satisfied as soon as the following conditions on the loss $c(r, \theta)$ are met:
\begin{enumerate}[(a)]
\item it is point additive,
\item it is convex with respect to its parameter $\theta$,
\item it can be stored and updated efficiently.
\end{enumerate}
\end{proposition}

It is possible to include an additional penalty term in the loss function.
For example, in the case of RNA-seq data one could add a lasso ($\lambda |\theta|$) or ridge penalty ($\lambda \theta^2$) to encode that 
a priori the coverage in most regions should be close to 0.
Our C++ implementation of the pruned DPA includes the possibility to add such a penalty term, however we do not provide an R interface to this functionality in our R package. 
One of the reason for this choice is that choosing an appropriate value for $\lambda$ is not a simple problem.

\section{Contribution}

\subsection{Pruned dynamic programming algorithm for count data}

We now show that the PDP algorithm can be applied to the segmentation of RNA-Seq data using a negative binomial model, and propose a criterion for the choice of $K$. Though not discussed here, our results also hold for the Poisson segmentation model.

\paragraph{Negative binomial model.}
We consider that in each segment $r$ all $y_t$ are the realization of random variables $Y_t$ which are independent and
follow the same negative binomial distribution. Assuming the
dispersion parameter $\phi$ to be known, we will use the natural
parametrization from the exponential family (also classically used in
\proglang{R}) so that parameter $\theta_r$ will be the probability of
success. In this framework, $\theta_r$ is specific to segment $r$ whereas
$\phi$ is common to all segments.

We have $E(Y_t)=\phi (1-\theta)/\theta$ and $Var(Y_t)=\phi
(1-\theta)/\theta^2$. We choose the loss as the negative
log-likelihood associated to data-point $t$ belonging to segment $r$
: $ -\phi \log (\theta_r) - y_t \log (1 - \theta_r) + A(\phi,y_t)$,
or more simply $\gamma(y_t, \theta_r) = -\phi \log(\theta_r) -
y_t \log( 1 - \theta_r) $ since $A$ is a function that does not
depend on $\theta_r$.

\paragraph{Validity of the pruned dynamic programming algorithm for the negative binomial model}
\begin{proposition}
Assuming parameter $\phi$ to be known, the negative binomial model satisfies (a), (b) and (c):
\end{proposition}

    \begin{enumerate}[(a)]
        \item As we assume that $Y_t$ are independent we indeed have that the loss is point additive : $ c(r, \theta) = \sum_{t \ \in \ r} \gamma(y_t, \theta).$
        \item  As $\gamma(y_t, \theta) = -\phi \log(\theta) - y_t \log( 1 - \theta) $ is convex with respect to $\theta$, $c(r, \theta)$ is also convex as the sum of convex functions.
        \item Finally, we have $c(r,\theta) = - n_r \phi \log(\theta) + \sum_{t \ \in \ r} y_t \log( 1 - \theta)$. This function can be stored and updated using only two doubles: one for $- n_r \phi$, and the other for $\sum_{t \ \in \ r} y_t$.
    \end{enumerate}

\paragraph{Estimation of the overdispersion parameter.}
We propose to estimate $\phi$ using a modified version of the estimator proposed by \cite{jonhson_kotz}:  compute the moment estimator of $\phi$ on each sliding window of size $h$ using the formulae $\phi=\mathbb{E}(Y)^2/(Var(Y)-\mathbb{E}(Y))$ and keep the median $\widehat{\phi}$. 

\subsection{C++ implementation of the pruned DPA}

We implemented the pruned DPA in C++ with in mind the possibility
of adding new loss functions in potential future applications. The
difficulties we had to come through were the versatility of the
program to design and the design of the objects it could work on.
Indeed, the use of full templates implied that we used stable sets of
objects for the operations that were to be performed on.

Namely: 

\begin{itemize}

\item The sets were to be chosen in a \emph{tribe}. This means that
  they all belong to a set ${\cal S}$ of sets such that any set
  $s\in{\cal S}$ can be conveniently handled and stored into the
  computer. A set of sets ${\cal S}$ is said \emph{acceptable} if it satisfies:
  
\begin{enumerate}
\item if $s$ belongs to $s$, $\mathbb{R} \setminus s\in{\cal S}$
\item if $s_1, s_2\in{\cal S},\,\, s_1\cap s_2 \in{\cal S}$
\item if $s_1, s_2\in{\cal S},\,\, s_1\cup s_2 \in{\cal S}$
\end{enumerate}

\item The cost functions were chosen in a set ${\cal F}$ such that

\begin{enumerate}
\item each function may be conveniently handled and stored by the software
\item for any $f\in{\cal F}$,  $f(x) = 0$ can be easily solved and the set of solutions belongs to an acceptable set of sets
\item for any $f\in{\cal F}$ and any constant $c$, $f(x) \leq c$ can be easily solved and the set of solutions belongs to an acceptable set of sets
\item for any $f, g \in {\cal F},\, f+g\in{\cal F}$.

\end{enumerate}

\end{itemize}

Thus we defined two collections for the sets of sets, intervals and parallelepipeds, and implemented the loss functions
corresponding to negative binomial, Poisson or normal distributions.  The
program is thus designed in a way that any user can add his own cost
function or acceptable set of probability function and use it without
rewriting a line in the code.

\subsection{Model Selection}

The last issue concerns the estimate of the number of segments $K$.
This model selection issue can be solved using penalized
$\log$-likelihood criterion where the choice of a good penalty
function is crucial. This kind of procedure requires the visit of
the optimal segmentations in $k=1,\ldots, K_{\max}$ segments where
$K_{\max}$ is generally chosen smaller than $n$. The most popular
criteria (AIC \cite{akaike} and BIC \cite{yao_estimation_1984})
failed in the segmentation context due to the discrete nature of the
segmentation parameter. In a non-asymptotic point of view and for
the negative binomial model, \cite{cleynen2013segmentation} proposed
to choose the number of segments as follows: denoting $\hat{m}_K$
the optimal segmentation of the data in $K$ segments,
\begin{eqnarray}
\hat{K}=\arg \min_{K \in 1:K_{\max}} \left\{\sum_{r \in \hat{m}_K} \sum_{t \in r}
\left [-\phi \log \dfrac{\phi}{\phi + \bar{y}_r} - Y_t
  \log(1-\dfrac{\phi}{\phi + \bar{y}_r}) \right ] +\beta K \left(1 +
4\sqrt{1.1+\log{\left(\frac{n}{K}\right)}}\right)^2 \right\},
\end{eqnarray}
where $\bar{y}_r = \dfrac{\sum_{t\in r} y_t}{\hat{n}_r}$ and
$\hat{n}_r$ is the size of segment $r$. The first term corresponds
to the cost of the optimal segmentation while the second is a
penalty term which depends on the dimension $K$ and of a constant
$\beta$ that has to be tuned according to the data (see the next section). With this choice of penalty,
so-called oracle penalty, the resulting estimator satisfies an
oracle-type inequality. A more complete performance study is done in
\cite{cleynen2013segmentation} and showed that the proposed
criterion outperforms the existing ones.

\subsection{R package}

The Pruned Dynamic Programming algorithm is available in the function
\code{Segmentor} of the \proglang{R} package
\pkg{Segmentor3IsBack}. The user can choose the distribution with the
slot \texttt{model} (1 for Poisson, 2 for Gaussian homoscedastic, 3
for negative binomial and 4 for segmentation of the variance). It
returns an S4 object of class Segmentor which can later be processed
for other purposes. The function \code{SelectModel} provides four
criteria for choosing the optimal number of segments: AIC
\cite{akaike}, BIC \cite{yao_estimation_1984}, the modified BIC
\cite{zhang_modified_2007} (available for Gaussian and Poisson
distribution) and oracle penalties (available for the Gaussian
distribution \cite{lebarbier_detecting_2005} and for the Poisson and
negative binomial \cite{cleynen2013segmentation} as described
previously). This latter kind of penalties require tuning a constant
according to the data, which is done using the slope heuristic
\cite{Arl_Mas-pente}.

Figure 4 (which is detailed in the Results and discussion section) was
obtained with the following $4$ lines of code (assuming the data was
contained in vector \texttt{x}):\newline

\texttt{> Seg<-Segmentor(x,model=3,Kmax=200)}

\texttt{> Kchoose<-SelectModel(Seg, penalty="oracle")}

\texttt{> plot(sqrt(x),col='dark red')}

\texttt{> abline(v=getBreaks(Seg)[Kchoose, 1:Kchoose],col='blue')} \newline

\begin{figure}[h]
 \begin{center}
 \subfigure[4 segments]{\includegraphics[width=7cm]{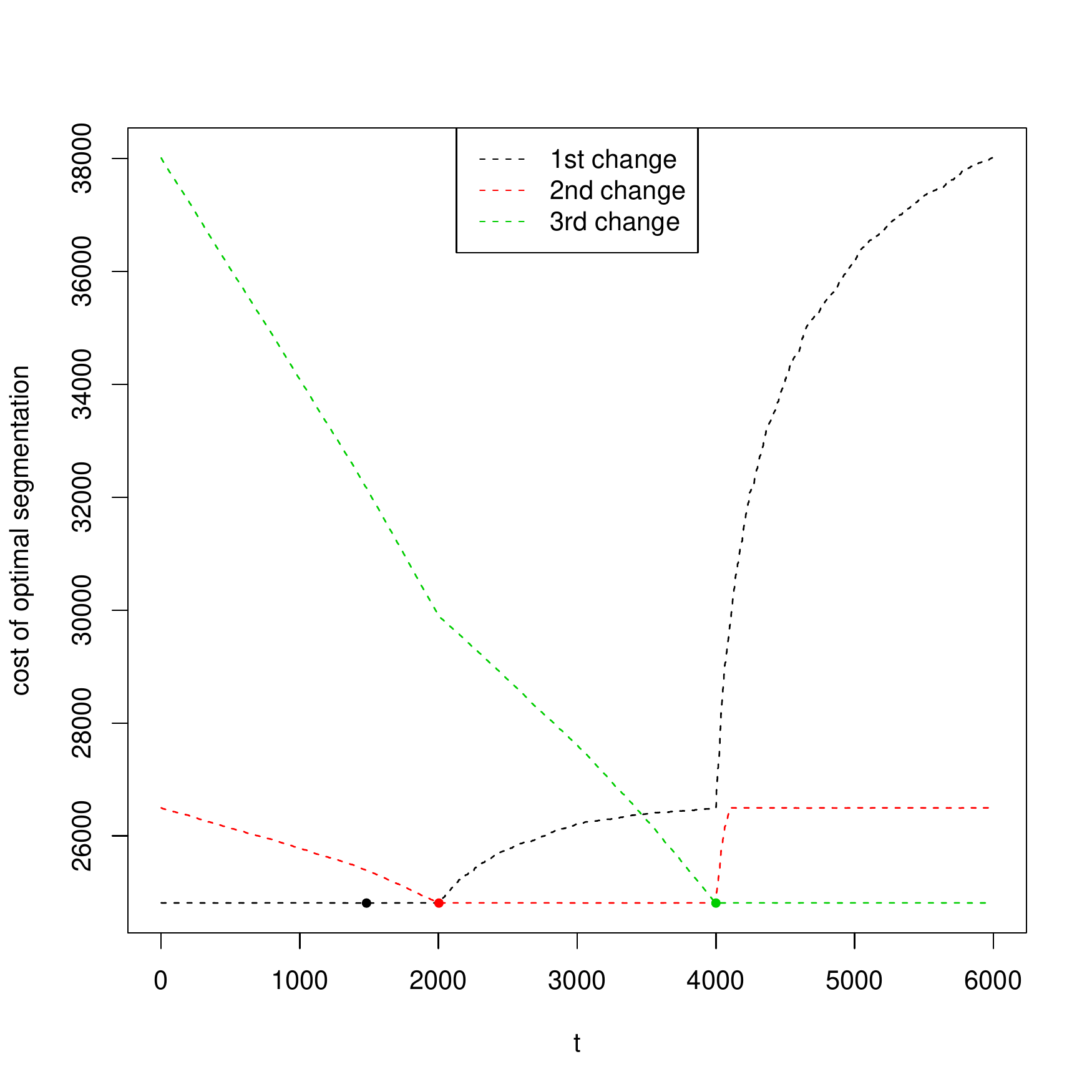}\label{real1}}
 \subfigure[3 segments]{\includegraphics[width=7cm]{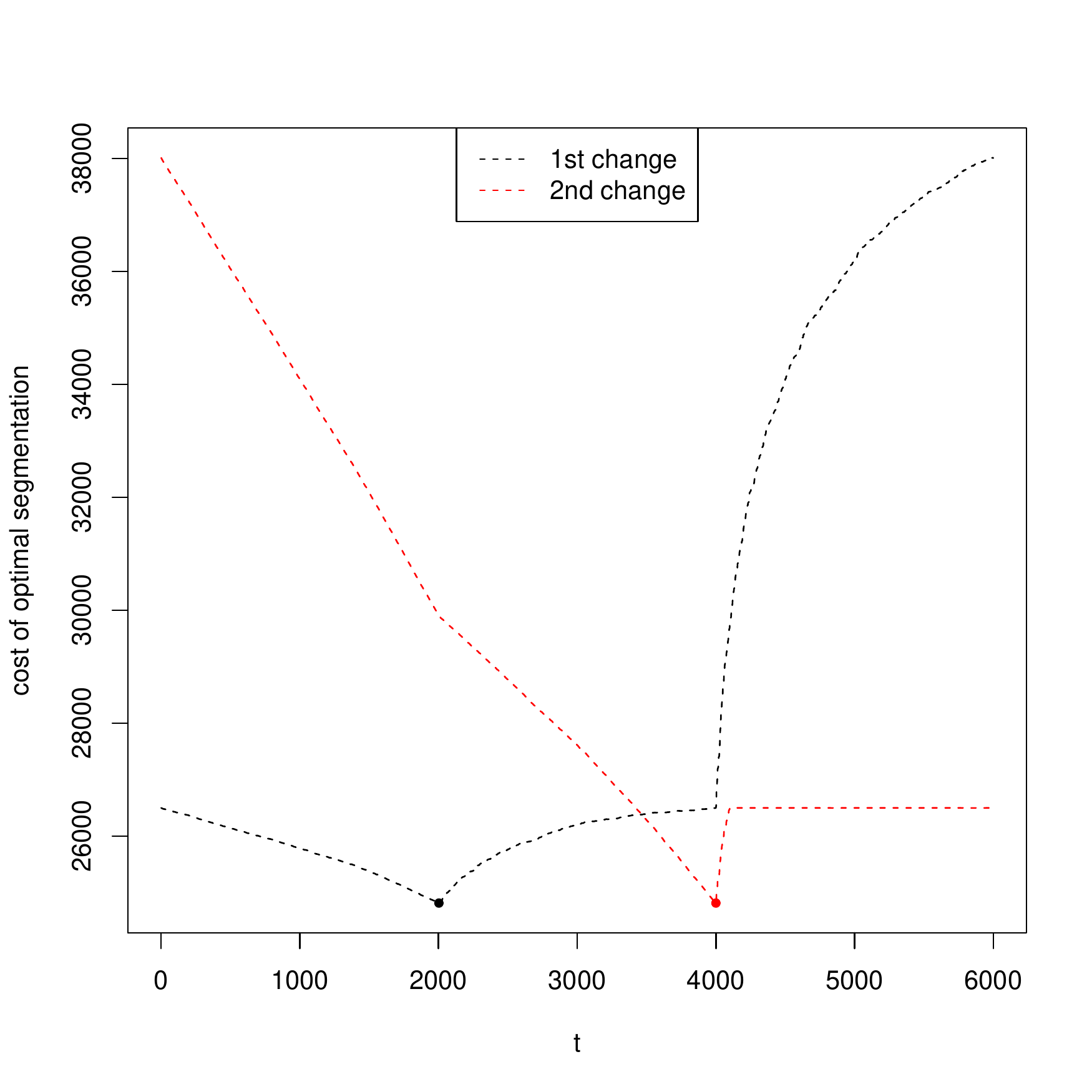}\label{real2}}
 \end{center}
 \caption{{\bf Cost of optimal segmentation in $4$ and $3$ segments.} Cost of optimal segmentation depending on the location of the $j^{th}$
change-point when the number of segments is $4$ (figure 1(a)) and $3$
(figure 1(b)) and the signal was simulated with $3$
segments. Illustration of the output of function
\texttt{BestSegmentation}.}\label{real-data}
\end{figure}

The function \code{BestSegmentation} allows, for a given $K$, to find
the optimal segmentation with a change-point at location $t$ (slot
\texttt{\$bestSeg}). It also provides, through the slot
\texttt{\$bestCost}, the cost of the optimal segmentation with $t$ for
$j^{th}$ change-point. Figure \ref{real2} illustrates this result for the
optimal segmentations in $4$ segments of a signal simulated with only
$3$ segments. We can see for instance that any choice of first
change-point location between $1$ and $2000$ yields almost the same
cost (the minimum is obtained for $t=1481$), thus the optimal
segmentation is not clearly better than the second or third. On the
contrary, the same function with $3$ segments shows that the optimal
segmentation outperforms all other segmentations in $3$ segments (Figure \ref{real1}).

\section{Results and discussion}

\subsection{Performance study}

We designed a simulation study on the negative binomial distribution
to assess the performance of the PDP algorithm in terms of computational
efficiency, while studying the impact of the overdispersion parameter
$\phi$ by comparing the results for two different values of this
parameter. After running different estimators (median on sliding
windows of maximum, quasi-maximum likelihood and moment estimators) on
several real RNA-Seq data (whole chromosome and genes of various
sizes) we fixed $\phi_1=0.3$ as a typical value for highly dispersed
data as observed in real RNA-Seq data, and chose $\phi_2=2.3$ for
comparison with a reasonably dispersed data-set.  For each value, we
simulated data-sets of size $n$ with various densities of number of
segments $K$, and only two possible values for the parameter $p_J$:
$0.8$ on even segments (corresponding to low signal) and $0.2$ on odd
segments for a higher signal. We had $n$ vary on a logarithmic scale
between $10^3$ and $10^6$ and $K$ between $\sqrt{n}/6$ and
$\sqrt{n}/3$. For each configuration, we segmented the signal up to
$K_{\max}=\sqrt{n}$ twice: once with the known value of $\phi$ and once
with our estimator $\widehat{\phi}$ as described above. 
We started with a window width $h = 15$. When the estimate was negative, we doubled $h$ and repeated the experience until the median is positive. 

Each configuration was simulated $100$ times.

For our analysis we checked the run-time on a standard laptop, and
assessed the quality of the segmentation using the Rand Index
$\mathcal{I}$. Specifically, let $C_t$ be the true index of the
segment to which base $t$ belongs and let $\hat{C}_t$ be the index
estimated by the method, then
$$ \mathcal{I}= \frac{2\sum_{t>s} \left[ \mathbf{1}_{C_t =
      C_s}\mathbf{1}_{\hat{C}_t = \hat{C}_s} +\mathbf{1}_{C_t \neq
      C_s}\mathbf{1}_{\hat{C}_t \neq
      \hat{C}_s}\right]}{(n-1)(n-2)}.
$$

\begin{figure}[!h]
 \centerline{\includegraphics[width=10cm]{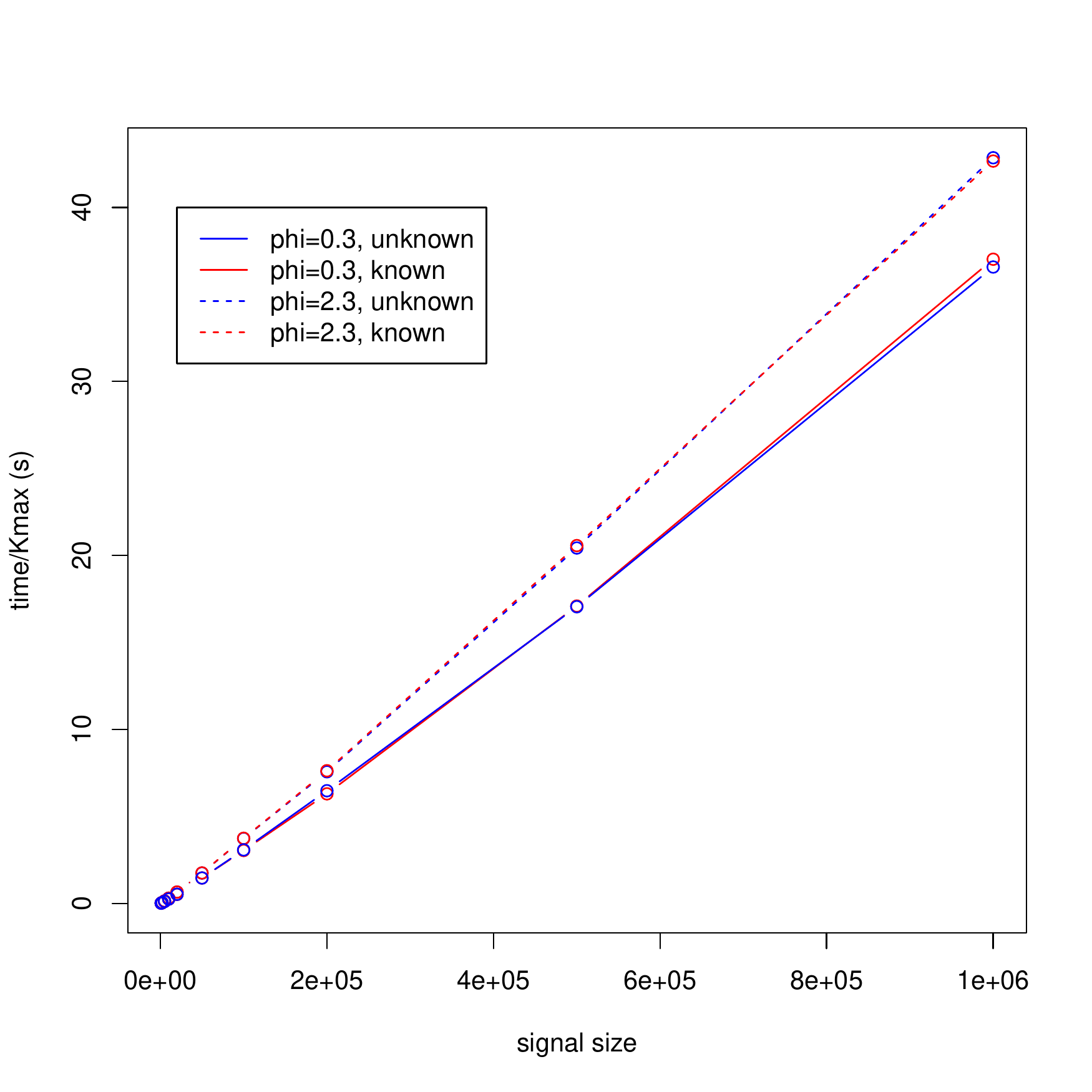}}
 \caption{{\bf Run-time analysis for segmentation with negative binomial distribution.}This figure displays the normalized (by $K_{\max}$) run-time in seconds
of the \pkg{Segmentor3IsBack} package for the segmentation of signals
with increasing length $n$, for two values of the dispersion $\phi$,
and with separate analysis when its value is known or estimated. While
the algorithm is faster for more over-dispersed data, the estimation
of the parameter does not slow the processing.}\label{fig2}
\end{figure}

\begin{figure}[!h]
 \centerline{\includegraphics[width=10cm]{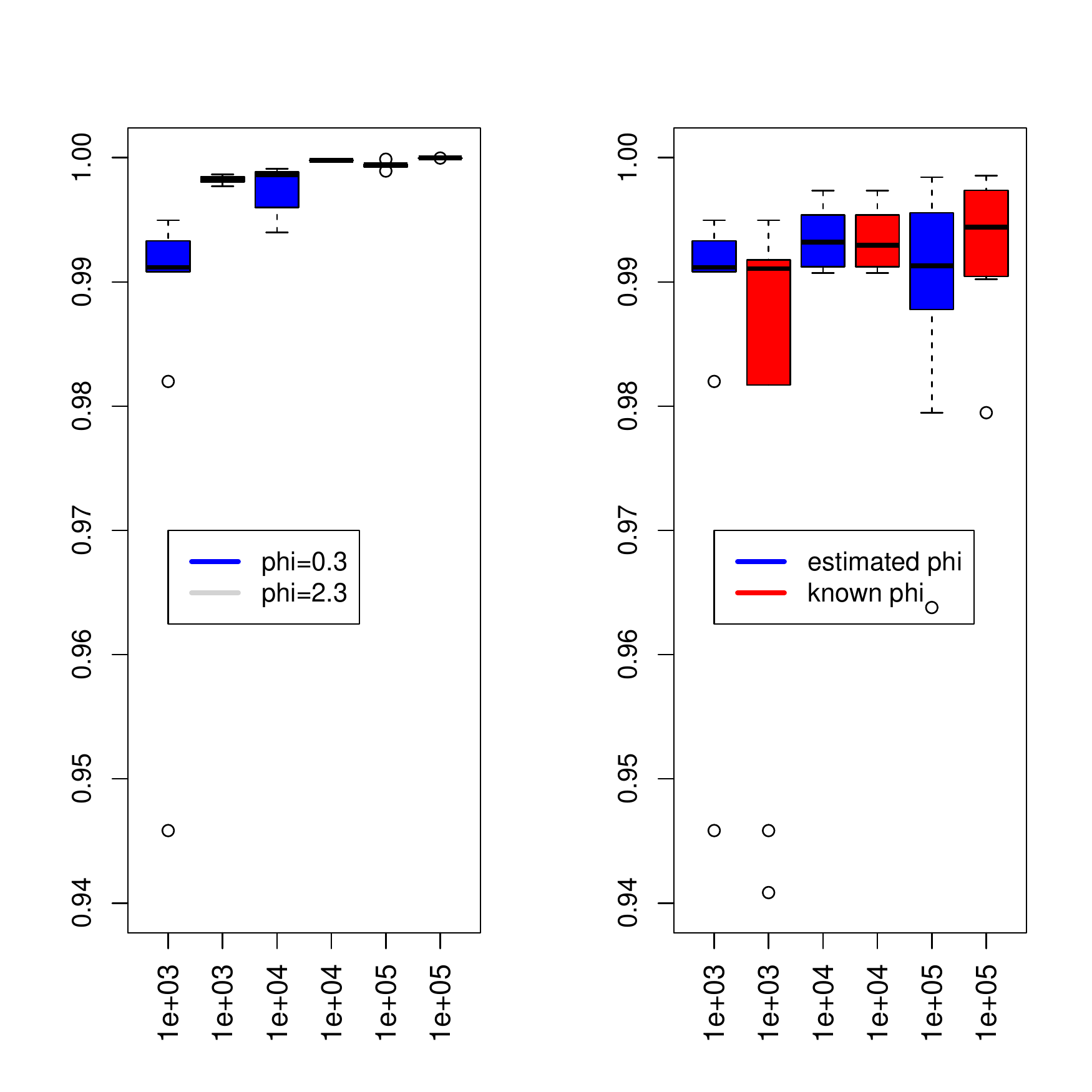}}
 \caption{{\bf Rand-Index for the quality of the segmentation.} This figure displays the boxplot of the Rand-index computed for each
of the hundred simulations performed in the following situations:
comparing the values with $\phi_1$ and $\phi_2$ when estimated (left
figure), and comparing the impact of estimating $\phi_1$ (right
figure). While the estimation does not decrease the quality of the
segmentation, the value of the dispersion affects the recovery of the
true change-points.}\label{fig3}
\end{figure}

Figure \ref{fig2} shows, for the particular case of $K=\sqrt{n}/3$, the almost
linear complexity of the algorithm in the size $n$ of the signal. As
the maximal number of segments $K_{\max}$ considered increased with
$n$, we normalized the run-time to allow comparison. This underlines
an empirical complexity smaller than $\mathcal{O}(K_{\max}n\log n)$,
and independent on the value of $\phi$ or its knowledge. Moreover, the
algorithm, and therefore the pruning, is faster when the
overdispersion is high, phenomenon already encountered with the $L^2$
loss when the distribution of the errors is Cauchy. However, the
knowledge of $\phi$ does not affect the run-time of the
algorithm. Figure \ref{fig3} illustrates through the Rand Index the quality of
the proposed segmentation for a few values of $n$. Even though the
indexes are slightly lower for $\phi_1$ than for $\phi_2$ (see left
panel), they range between $0.94$ and $1$ showing a great quality
in the results. Moreover, the knowledge of $\phi$ does not increase
the quality (see right panel), which validates the use of our
estimator.

\subsection{Yeast RNAseq experiment}

\begin{figure}[!h]
\begin{center}
\includegraphics[width=14cm]{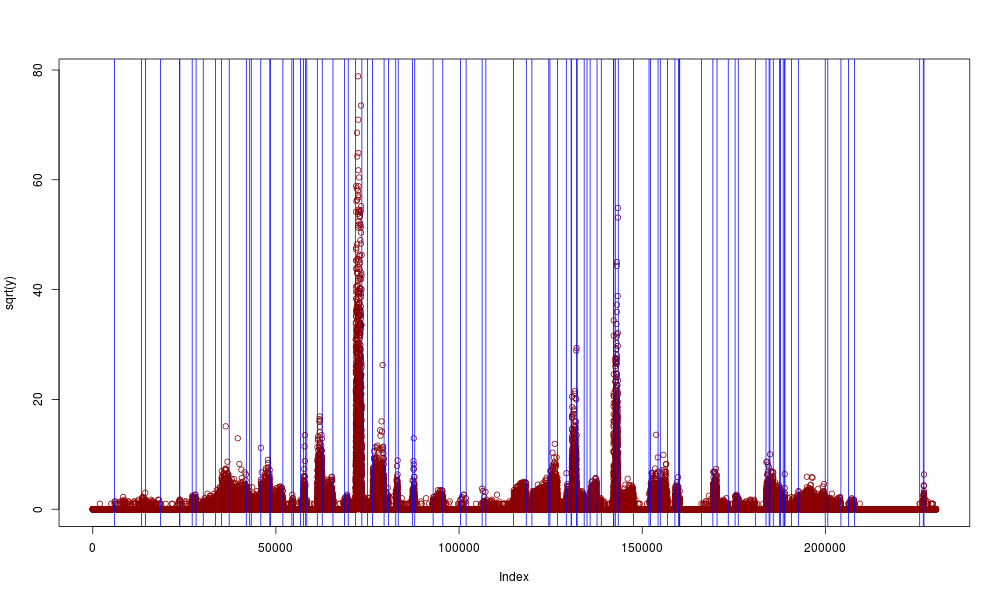}
\end{center}
\caption{{\bf Segmentation of the yeast chromosome 1 using the negative binomial loss.} The model selection procedure chooses $K=103$ segments, most of which
surround genes given by the SGD annotation.}\label{fig4}
\end{figure}

We applied our algorithm to the segmentation of chromosome $1$ of the
S. Cerevisiae (yeast) using RNA-Seq data from the Sherlock Laboratory
at Stanford University \cite{Risso_norma} and publicly available from
the NCBI's Sequence Read Archive (SRA,
\url{http://www.ncbi.nlm.nih.gov/sra}, accession number SRA048710). We
selected the number of segments using our oracle penalty described in
the previous section.  An existing annotation is available on the
Saccharomyces Genome Database (SGD) at
\url{http://www.yeastgenome.org}, which allows us to validate our
results. \\
With a run-time of $25$ minutes (for a signal length of $230218$), we
selected $103$ segments with the negative binomial distribution, most of
which (all but $3$) were found to surround known genes from the SGD.  Figure \ref{fig4}
illustrates the result.

\section{Conclusion}

Segmentation has been a useful tool for the analysis of biological
data-sets for a few decades. We propose to extend its application with
the use of the Pruned Dynamic Programming algorithm for count
data-sets such as outputs of sequencing experiments. We show that the
negative binomial distribution can be used to model such data-sets on
the condition that the overdispersion parameter is known, and proposed
an estimator of this parameter that performs well in our segmentation
framework.

We propose to choose the number of segments using our oracle
penalty criterion, which makes the package fully operational. This package also allows the use
of other criteria such as AIC or BIC. Similarly, the algorithm is not
restricted to the negative binomial distribution but also allows the
use of Poisson and Gaussian losses for instance, and could easily be
adapted to other convex one-parameter losses.

With its empirical complexity of $\mathcal{O}(K_{\max}n\log n)$, it can be
applied to large signals such as read-alignment of whole chromosomes,
and we illustrated its result on a real-data sets from the yeast
genomes. Moreover, this algorithm can be used as a base for further
analysis. For example, \cite{luong2012fast} use it to initialize their
Hidden Markov Model to compute change-point location probabilities.

 \bibliographystyle{plainnat}  
  \bibliography{Biblio}      

\end{document}